# Triage strategies for agile core sorting in extreme value scenarios

**Saeed Z. Gavidel**[1§], **J. L. Rickli**[1]

[1]Department of Industrial and Systems Engineering, Wayne State University, 4815 Fourth Street, Detroit, MI, USA
[§]Corresponding author

Email addresses:
SZG:  szgavidel@wayne.edu
JLR:  jlrickli@wayne.edu

## Abstract

Surveys have indicated that the remanufacturing industry is concerned about the necessity of agile and prioritized core sorting due to its potential benefits to optimal core inventory and condition assessment, both at equipment and component levels. As such, core sorting holds a pivotal role in generalized remanufacturing operations, however, extreme value core arrivals, its stochastic nature and resulting sorting issues warrant targeted modelling and analysis. This paper is devoted to triage as an agile and quality-based sorting strategy in extreme value scenarios, which can be utilized as a complementary core sorting strategy. A statistical model of extreme core arrivals is developed based on Extreme Value (EV) theory and related Generalized Extreme Value (GEV) and Fréchet (Fisher-Tippett type-II) distributions. The model is applied to extreme arrivals of valves in an industrial valve repair shop in order to formulate extreme arrival sorting strategies. Using a large sample size, distribution parameters are estimated and the stochastic behaviour of the extreme valve arrivals is evaluated and verified. A generic analogy between medical triage and remanufacturing triage is discussed, because triage can be used to address extreme core arrivals, associated statistical distributions, and their effect on core condition assessment in order to enhance management of unpredictable consequences of extreme core arrivals.

Keywords: Generalized Extreme Value (GEV); Triage; Remanufacturing; Sorting

## Introduction

Typically, upon arrival, cores are sorted according to their quality level [1]. Sorting operations are of great concern for cost of quality (COQ) activities [2], due uncertain core quality, quantity, and return timing. Acquired core condition often has high quality variability, which, along with the likelihood of non-remanufacturability of some cores, imposes prioritization in core sorting processes. In some circumstances, such as extreme arrivals of remanufacturing cores, due to scarcity of available operational and time resources, agile and even inaccurate sorting, *i.e.* triage, is beneficial under operational, timing, inventory, and market requirements.

As such, extreme value core arrivals, its stochastic nature and resulting sorting issues warrant targeted modelling and analysis. The main motivation of this research is modelling of less-expected but extreme value core arrivals to remanufacturing facilities and proposing triage as an operational tool and strategy to manage these rare but critical situations. Remanufacturing triage can be used to prevent adverse dimensions of unpreparedness such as business sluggishness in highly demanding and competitive market atmosphere due to lack of accessible resources. Loss of profit due to late supply of remanufactured products to market, customer defection, and other unpredictable penalties are some common consequences of inadequate response of remanufacturing process that will impact remanufacturing business system. Common strategies for prediction, planning, scheduling, and timing of remanufacturing operations for core arrivals that are based on traditional central tendency approaches such as modelling of arrivals by Poisson or Normal distributions are not sufficient in cases of extremely large or extremely small arrivals of cores. These scenarios can be considered as shock waves to a remanufacturing facility and consequently to its business system.

In this paper, the stochastic behaviour of remanufacturing systems under extreme situations is investigated and a statistical model is developed and applied on a real industrial case. Triage and its methods are introduced as tools to manage extreme value situations in remanufacturing facilities. Triage is a typical approach in medicine to address extremal situations, so, a generic analogy between medical triage and remanufacturing triage is described. The triage sorting strategy is based on statistical models of extreme core arrivals based on Extreme Value (EV) theory and related Generalized Extreme Value (GEV) and Fréchet (Fisher-Tippett type-II) distributions. The validity of the model is tested by hypothesis testing and comparisons between common, central tendency and less-likely but adverse extremal tendency are made. Lastly, it is suggested that the EVA approach in remanufacturing sorting issues should be considered as a complementary sorting strategy to central tendency approaches in order to have a sufficient level of preparedness for extremal situations.

**Triage as a Sorting Strategy**

Remanufacturing triage is an agile and less accurate core sorting strategy. According to current studies and reports, triage is used in some remanufacturing facilities as a core inspection strategy [3], as a part of their main sorting strategy for individual or bulk return that can be as much as 30,000 returns per week [4], and as an inaccurate but quick and profitable sorting method when the sorting, disposal, and transportation costs are reasonable in comparison to disassembly and reprocessing costs [5]. Remanufacturing triage and medical triage have some common logic resemblances. In medicine, triage, is a common patient prioritization sorting technique based on gravity of health condition and availability of medical resources like, operational resources, manpower, and time. This agile sorting and prioritization technique is of special significance when the normal balance between existing resources and the number of patient arrivals is unbalanced.

Medical triage can defined as, "the process of determining the priority of patients' treatments based on the severity of their condition. This rations patient treatment efficiently when resources are insufficient for all to be treated immediately." Historically, medical triage has been developed based on various approaches [6,7]. Table 1 summarizes common medical triage approaches and the proposed analogies for remanufacturing core sorting. The concept of medical triage began in the armed forces [6] where medical demands were greater than available resources, this situation in remanufacturing is equivalent to any situation that disturbs the normally expected stream of cores and creates an extreme amount of core flow to remanufacturing facility, so that the arrival rate and core handling capacity of the remanufacturing facility will have exorbitant gaps.

**Table 1 - Proposed analogy between medical triage and remanufacturing triage**

| | Medicine | Proposed Remanufacturing Equivalence | Prioritizing Factor |
|---|---|---|---|
| **Traditional** | Little or no formal effort to provide care | No prioritized core sortation | No criteria |
| **Baron Dominique-Jean Larry** | Recognition, evaluation and categorization, Treatment of the most urgent cases regardless of ranking and deferring less or fatally wounded cases | Condition assessment and processing, the most in demand cores have top priority | Market pull factor |
| **Wilson** | Concentration of treatment on most likely successful cases, Some low chance cases will die | Concentrated remanufacturing on cores with high remanufacturability No chance of second life for less remanufacturable cores | Likelihood of success |
| **First Come First Served (FCFS)** | Treatment based on order of arrival and regardless of gravity of wounds, rank or any other criteria | Remanufacturing just based on core arrival order without any other criteria and means equal chance for all for second life | Order of arrival |
| **Great Goodness for Greatest Number (GGGN)** | Depriving of urgent cases needing huge amount of attention, time and resources in favor of dozens of other cases | Remanufacturing dozens of cores with high remanufacturibiltiy instead of one needing great amount of resources | Number of treatment |
| **Less Severity First Treatment (LSFT)** | Prioritizing less urgent cases since they can be treated swiftly and can return to action | Prioritization of cores with less defects regardless of other criteria | Less defects |
| **Maximize the Fighting Strength (MFS)** | Most quick return to duty with least expenditure of time and resources | Swift response to market demand and minimization of remanufacturing time, customer urgent requirements | Time |

One of the goals of sorting of cores based on triage is providing a complementary sorting tool to be utilized along with facilities' main sorting strategy, such as central tendency sorting approaches. The use of triage sorting is contingent upon the extreme arrivals of cores that can be interpreted as severe unbalance between core arrivals and facility remanufacturing capacity. Hence, considering triage as a complementary strategy of core sorting management may be of interest for some facilities, especially those having unexpected, extreme core arrivals. The flowchart in Fig. 1 shows a typical core reverse supply chain in remanufacturing industry under normal operations, but this status can be disrupted by an extreme shock wave of arrivals.

Since agile core sorting by triage approach is contingent upon extreme situations, there is close relation between triage and extreme value scenarios. Hence, modelling and analysis of extreme value scenarios is an essential prerequisite for triage analysis. The next section addresses extreme value scenario modelling and analysis and introduce a sample example from a simulated data set will be introduced to illustrate the need for extreme value analysis. It will be inferred that utilization of the central tendency approach in extreme value scenarios will lead to misinterpretations. Subsequently, if a decision is taken in association with EVA scenarios, decision makers must take more conservative strategies to be protected against consequences of misinterpretation of system stochastic behaviour.

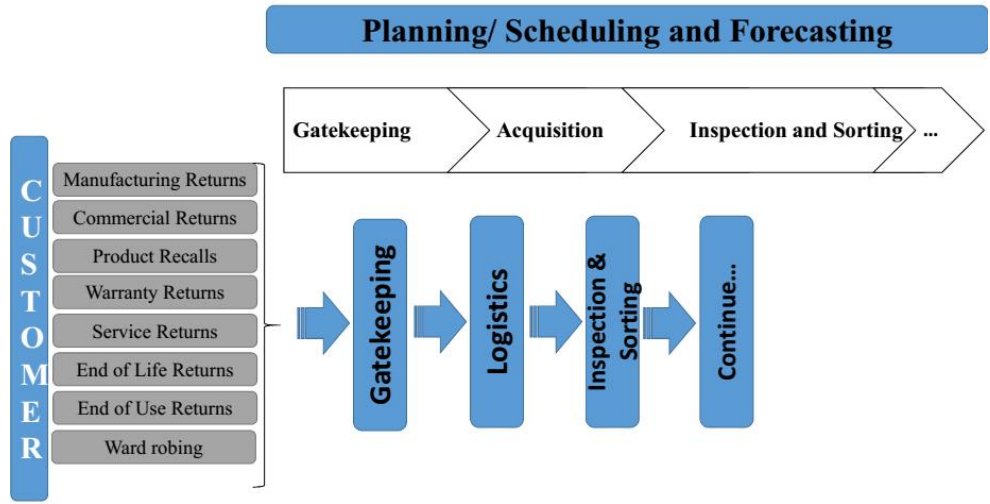

**Figure 1 - Three types of GEV distributions**

# Methods

Extreme Value Analysis (EVA) is a branch of statistics that deals with extremes, maxima, and minima of random variables.. In most investigations in remanufacturing, the focus is on central tendency approaches to deal with remanufacturing core management. Galbreth and Blackburn used normally distributed demand patterns for modelling optimum acquisition quantities [8], Aras et al. modelled both customer demand and product returns as Poisson Process for hybrid manufacturing and remanufacturing systems [9], and Teunter and Flapper assumed normally distributed behaviour for demands to model optimum core acquisitions [10]. In various engineering disciplines, EVA has applications to predict less-likely but still probable events with severe impacts, such as flood prediction in hydrology engineering, earthquakes for structural engineering, fatigue strength in extreme mechanical loading, and extreme temperature variations in meteorology [11, 12]. Each of these instances is concerned with unusual, extreme value behaviour and not average behaviours. Other applications in the field of strength of material, electrical engineering, highway traffic modelling, corrosion resistance, and pollution studies have also been reported [13].

Although extreme arrivals of remanufacturing cores are not frequent, they still have some probability of occurrence, and in the case of ignoring such events, remanufacturing systems may incur significant losses.

In this research, the nature and behaviour of extreme value core arrivals are studied based on Extreme Value Analysis and a statistical model is developed. Classical EVA theory is presented in many Statistics resources [11,12,13]. The foundation of EVA theory is the study of the statistical nature and behaviour of maxima, shown below as:

$$M_n = \max\{X_1, X_2, \dots, X_n\}$$

Where $\{X_1, X_2, \dots, X_n\}$, is a set of randomly distributed variables with common distribution function $F$. It should be noted that the distribution must be common but *not of a specific type*. Since $M_n$ is the maximum of the observed data, the distribution of $M_n$, confirms:

$$P(M_n < z) = P(X_1 \leq z, \ldots, X_n \leq z) = P(X_1 \leq z) \ldots P(X_n \leq z) = (F(z))^n$$

Since $F$ is the parent data distribution, it is necessary to estimate $F$ from an available set of sample data in order to study $M_n$. One common approach to estimate $(F(z))^n$, is use of extreme data. This idea is similar to that used in estimation procedure of the sample mean average. It is noteworthy that, since $F(z) < 1$, then for $z < z_{sup}$, where $z_{sup}$, is the smallest value of z such that, $F(z) = 1$, $F^n(z) \xrightarrow{n \to \infty} 0$, this issue is resolved by defining another variable such as, $M_n^* = \frac{M_n - b_n}{a_n}$, where $\{a_n\}, \{b_n\}$, are sequences of constants with $a_n > 0$. Here a well-established theorem known as *Extremal types* theorem is presented without proof.

**Theorem**: If there exist sequences of constants $\{a_n\}$ and $\{b_n\}$ such that

$$P\left\{\frac{M_n - b_n}{a_n} \leq z\right\} \xrightarrow{n \to \infty} G(z)$$

Where $G(z)$ is a non-degenerate distribution function, then $G(z)$ will belong to one of the following distribution types (Table 2), for more details refer to [11,12]. Table 2, presents three types of extremes value distributions with corresponding domains of random variable and names.

**Table 2 - Three types of extreme value distributions**

| Extremals | Type | Name |
|---|---|---|
| $G(z) = \exp\left\{-\exp\left(-\left(\frac{z-b}{a}\right)\right)\right\}, -\infty < z < +\infty$ | I | Gumble |
| $G(z) = \begin{cases} 0 & ; z \leq b \\ \exp\left\{-\left(\frac{z-b}{a}\right)^{-\alpha}\right\} & ; z > b \end{cases}$ | II | Fréchet (Fisher – Tippet) |
| $G(z) = \begin{cases} \exp\left\{-\left(-\left(\frac{z-b}{a}\right)^{\alpha}\right)\right\} & ; z < b \\ 1 & ; z > b \end{cases}$ | III | Weibull |

In all distributions, *a* is positive and *b* is real. In the second and third distributions $\alpha > 0$. The presented three distributions belong to one family of distributions called *Extreme Value distributions*. These three distributions are the only possible limits for the distribution of normalized extremes (maxima or minima) and this fact does not depend on the distribution, *F*, of parent data set.

A common approach in EVA theory is combining the three forms of the family in one simple form, called the *Generalized Extreme Value*, GEV distribution and is presented as follows:

$$G(z) = \exp\left\{-\left(1 + \xi\left(\frac{z-\mu}{\sigma}\right)\right)^{-1/\xi}\right\} \quad ; z > b$$

In the generalized format of extreme value distribution, $\xi\left(\frac{z-\mu}{\sigma}\right) > 0$ and $\xi, \mu, \sigma$, are three parameters of the GEV distribution. These three parameters and associated domains are summarized in table 3.

**Table 3 - The three parameters of the GEV distribution**

| Parameter | Symbol | Domain |
|---|---|---|
| *Location* | μ | *(-∞, +∞)* |
| *Scale* | σ | *(0, +∞)* |
| *shape* | ξ | *(-∞, +∞)* |

GEV converts to Fréchet distribution if the shape factor is positive, and converts to the Weibull distribution if it is negative. For Gumble type, shape factor is approaching zero. The basic advantage of unified GEV format vs. three individual distribution format is the simplification of statistical analysis. Figure 2, illustrates the three types of GEV distributions, Type-I, Gumble, has no upper or lower limiting values, Weibull has just upper limiting value with no lower limiting value, and Fréchet type has lower limiting value and no limiting value in right hand side, respectively.

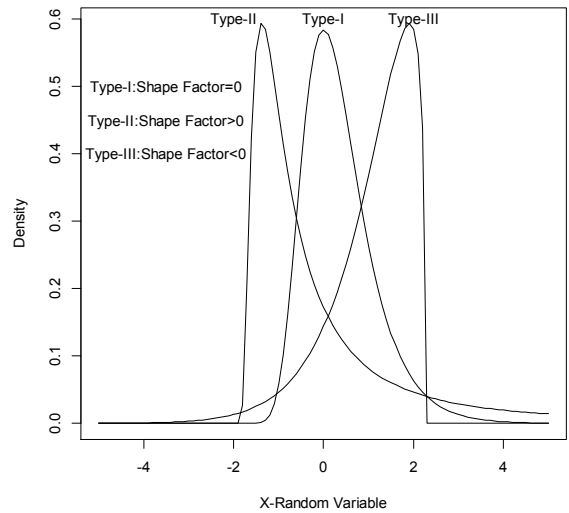

**Figure 2 – The three types of GEV distributions**

In order to apply the above results to a set of independent, identically distributed, *iid*, random variables like $X_1, X_2, \ldots, X_n$, the first step is *blocking* the data set into *n* blocks of observations, provided that *n* should be sufficiently large. Then *maxima* $z_i$ of each block is screened and eventually the appropriate GEV distribution can be fitted to the maxima $\{z_1, z_2, \ldots, z_i\}$. The estimation of distribution parameters in most cases is done by maximum likelihood estimator method, MLE, but other methods can be used, see [11,12], in this research due to popularity, the estimation of parameters are conducted by MLE. At this stage and prior to analysis of real industrial case, the model is used to analyse a simulated data set. The approach is applied to simulated hurricanes data generated by the US Information Technology Laboratory [14], data from Milepost 2700. Cumulative distribution functions for both EVA and central tendency approaches are presented in Fig. 3.

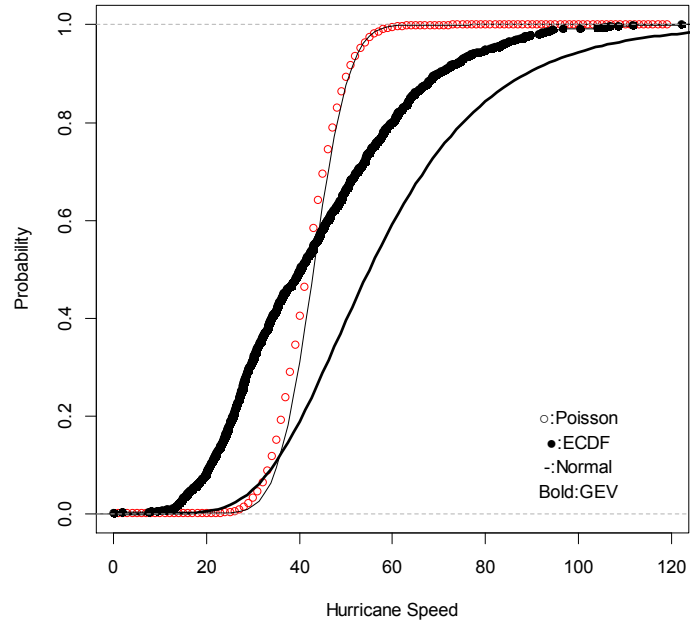

**Figure 3 - CDFs of simulated hurricane speeds for Milepost 2700**

As it can be observed from Fig. 3, the normal distribution predicts *zero* probability for occurrences above 80 units of wind speed. However, an examination of the data set reveals that there are 35 cases out of 1000 that have hurricane speeds of more than 80 and less than 100 units of speed, and the total number of exceeding over 80 is 43 out of 1000. Contrary to indications based on the normal and Poisson distributions, the probability of occurrence of extreme wind speeds is not zero [14].

## Results

The proposed triage approach using EVA is illustrated with a case study of industrial valve repair/remanufacturing operations. Industrial valves are mechanical devices usually used in connection with pressurized vessels to completely stop or regulate flow. Safety and relief valves, steam traps, and control valves are three kinds of industrial valves that are widely used in chemical complexes. Safety valves are vital for the protection of people and physical assets and steam traps have the duty of discharging condensate and non-condensable gases from piping system [15]. Valve maintenance and repair/remanufacturing is of critical importance from both safety and economic perspectives. Repair is an equivalent term used for remanufacturing in valve industry [16].

Arrivals of industrial valves to the valve shop of a chemical complex for a period of 476 consecutive operational days is investigated. Data collection was conducted based on formally issued organizational work orders through a Computerized Maintenance Management System (CMMS). The necessity of valve remanufacturing and their arrivals to valve shop in this industrial complex is determined based on various dominating factors such as results of sophisticated condition assessments, calendar-based Preventive Maintenance (PM), premature failures such as jamming and leakage, requests or requirements of operational units, partial or full overhauls, Opportunity-based Maintenance (OM), and safety and economic considerations required by management. It should be noted that the approximate mean remanufacturing capacity of valve shop is 10 units of valves per operational day. The investigation period includes all major arrivals listed in Table 4 except full overhauls. There are some arrivals of small quantity, like arrivals due to PM plans, premature failures, operational requirements, OM requests, and some small revamps. These small arrivals may be expected or unexpected. In this complex like many others, traditional condition assessment of steam traps is conducted by unit operators based on their experience and human senses. Hence, degrees of human error factors are expected. To have more reliable assessments and due to energy management purposes in steam network of the chemical complex, condition assessment of steam traps was outsourced. More reliable ultrasonic condition assessments resulted in elimination of human errors and huge arrivals were observed. Industrial disasters are usually followed by management enforcements to take more efficient preventive actions that in turn, cause increased levels of extreme safety and relief valve arrivals. Assumptions made is this case study are as follows; it is assumed that inspection and triaging processes are error proof, quality of all arrived cores are assumed to be equal and at levels requiring complete disassembly and testing, quantity of arrivals is considered as the sole uncertainty parameter and effects of other parameters like timing and detailed quality levels of cores are not considered.

The investigation period is divided to 158 three-day sub-periods and every 4 sub-periods are considered as one block. Maxima of these blocks are extracted as block maxima. Hence, there are 39 blocks and a half block, here for simplicity, the half block is treated as a full block. Scatter plot maximum values of blocks is depicted in Figure 4. Extreme values observed in this scatter plot are attributable to reasons such as, post-disaster preventive actions, partial overhauls, outsourced condition assessments, or a combination of factors.

**Table 4 - Summary of factors affecting valve arrivals**

| Arrival Determining Factor | Nature of Core Arrival | Variable Type | Amount |
|---|---|---|---|
| Preventive Maintenance | Preplanned/prescheduled | Deterministic | Usually small |
| Outsourced Condition Assessments | Preplanned/prescheduled | Stochastic | Extremely Huge |
| Premature Failures | Unexpected | Stochastic | Usually small |
| Operational Units | Unexpected | Stochastic | Usually small |
| Partial/Full Overhaul | Preplanned/prescheduled | Stochastic | Extremely Huge |
| Opportunistic Maintenance | Unexpected | Stochastic | Usually small |
| Post-disaster Preventive Actions | Unexpected | Stochastic | Extremely Huge |
| Revamps | Preplanned/prescheduled | Deterministic | Small or Huge |

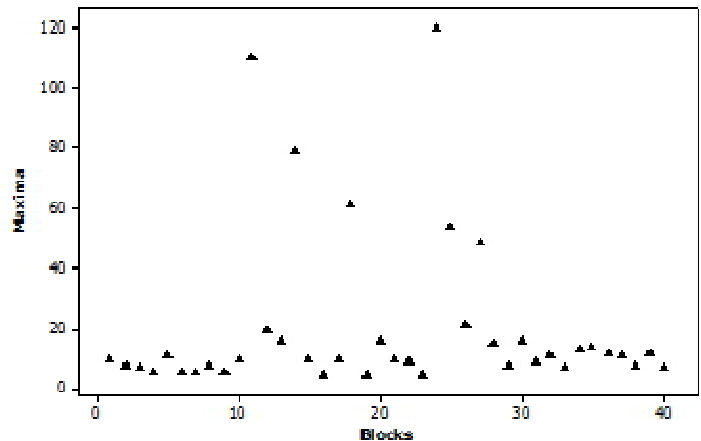

**Figure 4 - Scatter plot of maxima valve arrivals vs. blocks**

Table 5 presents the descriptive statistics for parent and block maxima data sets. Note that in EVA, the type of distribution for parent sample, from which the maxima blocks are generated is not of great concern. Figure 5, shows the histogram of blocks maxima.

**Table 5 - Descriptive statistics for Weekly and Biweekly core arrivals**

| Variable | N | Mean | SE Mean | St Dev. | Mode | Min. | Q1 | Med. | Q3 | Max. |
|---|---|---|---|---|---|---|---|---|---|---|
| Parent Sample | 158 | 11.84 | 1.39 | 17.44 | 5 | 0 | 5 | 7 | 10 | 120 |
| Block Maxima | 40 | 20.55 | 4.32 | 27.30 | 10 | 5 | 7.25 | 10 | 16 | 120 |

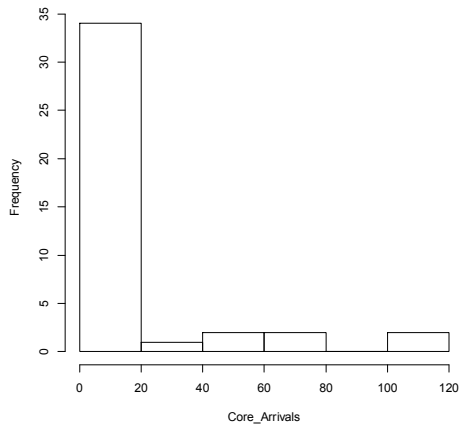

**Figure 5 - Histogram of maxima block valve arrivals**

Extracted maxima of blocks were used to estimate three parameters of GEV distribution. Table 6 presents the estimation results yielded by Maximum Likelihood Method (MLE). The corresponding confidence intervals are included. For the purpose of comparison, results coming from Probability Weighted Method (PWM) are also presented.

**Table 6 - Estimation of distribution parameters**

| Parameter | Symbol | PWM | MLE | SE | 95% CI |
|---|---|---|---|---|---|
| *location* | $\mu$ | 5.5583 | 8.3540 | 0.7565 | (6.8104,9.8976) |
| *scale* | $\sigma$ | 3.2649 | 4.2832 | 0.9220 | (2.7535,6.6628) |
| *shape* | $\xi$ | 0.5814 | 0.8903 | 0.2038 | (0.4762,1.3045) |

Figure 6 shows the cumulative distribution function plot for GEV, Experimental Cumulative Distribution Function (ECDF), and two central tendency distributions, normal and Poisson. Distribution parameters for both central tendency distributions were estimated from parent data set.

The conformance between the ECDF generated from actual data set and the GEV model, can be seen. This is critical in larger values of arrivals that is the main motivation of using GEV. Further investigations reveal that in the case of non-extreme, usual arrivals, there is an observed conformance between Poisson and ECDF distributions. It can be inferred from this figure, for arrivals greater than approximately 20 for Poisson and greater 40 for Normal distributions, the corresponding CDF's yield zero probability of occurrence while CDF of GEV distribution yields 7.37% of occurrence probability. Investigation of the actual data set reveals that in 9 cases core arrivals are more than 40, which indicates that the observed probability is approximately 5.7%.

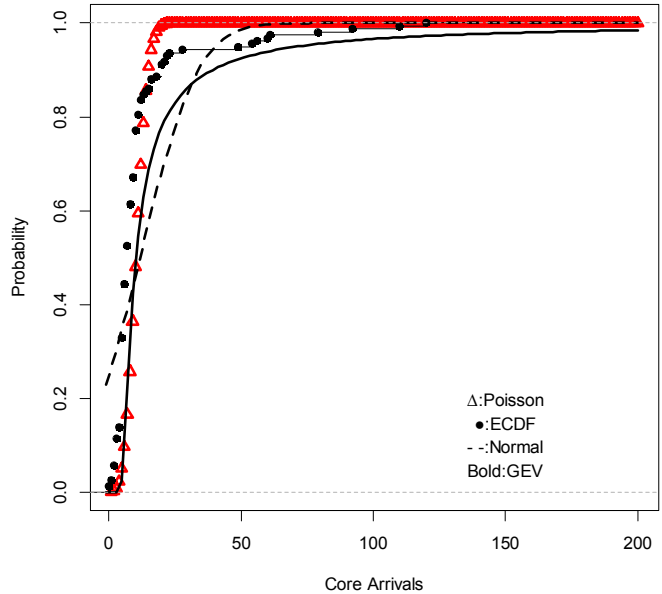

**Fig. 6 - CDFs for GEV, ECDF, normal, and Poisson approaches**

In the EVA approach, return level plays an important role. Return value is defined as a value that is expected to be equalled or exceeded on average once every interval of time (*T*-with a probability of *1/T*) [17]. Return level can be evaluated from the fitted distribution by estimating how often the extreme quantiles occur with a certain return level. Therefore, the CDF of the GEV distribution should be obtained and then be equalled to *(1-1/T)*. *T* is called return period and corresponding *z*-score is return value. The following figure shows the return period vs. return level plot for valve arrivals.

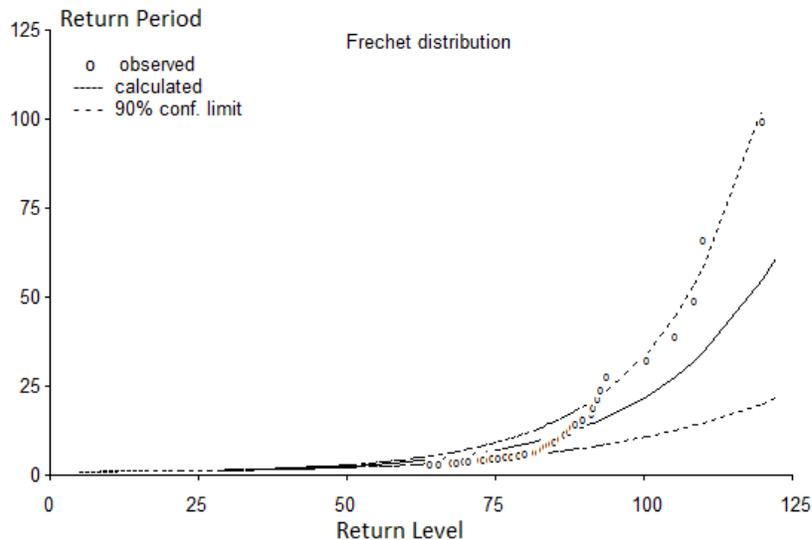

**Figure 7 - Return period vs. return level plot**

The developed model was verified by Pearson's Chi-squared Goodness Of Fit, GOF test, and Pearson's Chi-squared method [18]. The hypothesis testing was conducted by considering null hypothesis as Fréchet distribution versus the alternative that the distribution is not Fréchet. The predicted values resulting from the Fréchet distribution, observed data, and predicted values yielded by normal and Poisson distribution are presented in the Table 7. The *p-value* of this test was calculated and was equal to *0.2931*. This level of *p-value* is significantly greater than any common significance levels such as *0.01, 0.05 or 0.1*. Note that the *p-value* is considered to be as a measure of depth of confidence, and shows that this distribution conforms to GEV distribution type-II.

**Table 7 – Observed vs. Predicted arrivals for GEV and Normal scenarios**

| Bin | Observed | Predicted Probability GEV | Predicted Frequency GEV | Predicted Probability Normal | Predicted Frequency Normal | Predicted Probability Poisson | Predicted Frequency Poisson |
|---|---|---|---|---|---|---|---|
| 0-25 | 34 | 0.8298 | 33 | 0.7747 | 31 | 0.99987 | 40 |
| 0-50 | 35 | 0.9263 | 37 | 0.9876 | 40 | 1 | 40 |
| 0-75 | 37 | 0.9542 | 38 | 0.9999 | 40 | 1 | 40 |
| 0-100 | 37 | 0.9672 | 39 | 1 | 40 | 1 | 40 |
| 0-125 | 40 | 0.9746 | 39 | 1 | 40 | 1 | 40 |
| 0-150 | 40 | 0.9794 | 39 | 1 | 40 | 1 | 40 |
| 0-175 | 40 | 0.9827 | 39 | 1 | 40 | 1 | 40 |
| 0-200 | 40 | 0.9852 | 39 | 1 | 40 | 1 | 40 |

Like the hurricane sample case, the normal and Poisson distribution are not capable of efficiently predicting extreme cases, particularly in the case of 40 and more arrivals which predicts zero probability for more than 40 arrivals. However, the actual data indicates that there are 9 cases of 158 having more than 40 arrivals. The investigation of the actual data also suggests that even though the probability of extreme arrivals is quite low, it should not be discounted.

## Discussion

Agile and prioritized patient sorting has been conducted in medicine, but due to common logical similarities between medical triage and agile but perhaps less accurate core sorting in remanufacturing, medicine triage methods can be applicable to remanufacturing core sorting issues as suggested in Table 1. In remanufacturing, like other business systems, management of crisis is of great importance. Since extreme core arrivals and their stochastic nature impose extreme disturbances on remanufacturing sorting systems, proper management of extreme events is significant, especially from core sorting perspective. Results suggest that triage can be a critical soring approach that is complementary to primary sorting strategies. The stochastic behaviour of extreme value scenarios was investigated in real set of industrial data collected from remanufacturing of valves. As was expected and indicated by Fig. 6 and Table 7, predicting models developed by GEV are more efficient to predict stochastic behaviour of extreme core arrivals than common central tendency approaches. The expected return and period level of huge arrivals was obtained, which is important to have sufficient predictions of these two factors to have certain level of preparedness in remanufacturing systems to address disturbances efficiently. Results from hypothesis testing for goodness of fit for the model and actual data indicate that *p-value* of the test is 0.2931, which is satisfactorily greater than common significance levels used in both industry and theory. This indicates an efficient statistical model and, from a practical perspective, a more reliable model for predicting stochastic behaviour of systems and processes.

## Conclusions

The central tendency approach is not the sole approach to deal with core sorting in remanufacturing. In extreme value core arrivals, agile and even less accurate sorting strategies were suggested due to scarcity of available resources. In remanufacturing business systems, the common approach for modelling of core arrivals is primarily based on utilization of central tendency approaches such as Poisson or normal distributions that are shown to adequately characterize EVA scenarios. Hence, having targeted models for extreme core arrivals is critical to having complete sorting strategies. It was shown that the EVA approach is a powerful tool to evaluate such rare but highly impactful situations and assist decision makers.

As a contribution to remanufacturing in practical scope, triage (modelled with EVA) can be used as a complementary sorting approach to major sorting strategies. As mentioned, EVA approach assists decision makers to have stronger prediction abilities, especially for gatekeeping and core acquisition purposes, and this will protect remanufacturing facility from losses resulting from unpredicted extreme situations and associated consequences like overstocking, overproduction or losses due to sluggish response to market pull. The major contribution of this research is the application of EVA theory to remanufacturing sorting. EVA grants unique opportunities to open scopes to future researches such as investigation of multivariate timing and building equivalent statistical models and other triage sorting strategies for optimization of core acquisition in extreme value scenarios. Overall, results indicate that EVA is a powerful tool in dealing with extreme value scenarios in remanufacturing industry and utilization of this tool as a sorting approach or in combination with other sorting strategies can enhance remanufacturing operations.